\documentclass[prd,aps,amsfonts,eqsecnum,superscriptaddress,nofootinbib,longbibliography,notitlepage]{revtex4-1}

\usepackage{graphicx}
\usepackage{xcolor}
\usepackage{caption}
\usepackage{rotating}
\usepackage{amsmath,amssymb,graphics,amsthm,isomath}
\usepackage{subfigure}
\usepackage{float}
\usepackage{physics}

\usepackage[colorlinks=true, urlcolor=violet, linkcolor=blue, citecolor=red, hyperindex=true, linktocpage=true]{hyperref}
\usepackage[capitalise,compress]{cleveref}

\newtheorem{thm}{Theorem}
\numberwithin{thm}{section}

\newtheorem{lem}[thm]{Lemma}

\makeatletter
\renewcommand{\p@subsection}{}
\renewcommand{\p@subsubsection}{}
\makeatother

\usepackage{xcolor}
\usepackage{mathtools}

\usepackage{dsfont}

\begin{document}

\title{Fast high-fidelity multi-qubit state transfer with long-range interactions}

\author{Yifan Hong}
\email{yifan.hong@colorado.edu}
\affiliation{Department of Physics and Center for Theory of Quantum Matter, University of Colorado, Boulder CO 80309, USA}

\author{Andrew Lucas}
\email{andrew.j.lucas@colorado.edu}
\affiliation{Department of Physics and Center for Theory of Quantum Matter, University of Colorado, Boulder CO 80309, USA}

\begin{abstract}
We describe an efficient protocol to perform quantum state transfer using Hamiltonian dynamics with long-range interactions.  The time to transfer $n$ qubits a sufficiently large distance is proportional to $\sqrt{n}$.   Even without error correction, the fidelity of this multi-qubit state transfer process remains finite for arbitrarily well-separated qubits in the presence of uncorrelated random errors in coupling constants.
\end{abstract}

\date{\today}

\maketitle
\tableofcontents

\section{Introduction}
Rapid and high fidelity quantum state transfer is an important requirement for any practical large-scale quantum computer.  In a nutshell, suppose we have a set of $N$ qubits in the state \begin{equation}
    |\psi(0)\rangle = |\phi\rangle_i \otimes |0000\cdots\rangle_{-i}
\end{equation}
where $\phi$ is an arbitrary two-state wave function, and the $-i$  subscript means that all qubits except $i$ are initialized in the $|0\rangle$ state.  What is the smallest time $t$, evolving under some Hamiltonian $H(t)$, such that \begin{equation}
    |\psi(t)\rangle = |\phi\rangle_j \otimes |0000\cdots \rangle_{-j},
\end{equation}
i.e. how long would it take to transfer the full quantum state of qubit $i$ (including phase information) to qubit $j$?  This is, of course, entirely a question of the quantum hardware and/or architecture.  Even with the ability to apply only 2-local (two-body) Hamiltonians, we could clearly achieve state transfer in constant time if all pairwise couplings are allowed.  However, most near term realizations of a quantum device do not have all-to-all tunable couplings: in superconducting qubit arrays \cite{Boixo_2014} couplings are usually rather local in space, while in trapped ion crystals \cite{exp1} the couplings are all-to-all but not as tunable.  ``Designer graphs" like the $\lbrace 0,1\rbrace^N$ hypercube \cite{Christandl_2004}, where it is easy to perform perfect state transfer between qubits living on any two graph vertices, are not likely to be achieved in a near-term quantum information processor.

In the presence of inevitable noise in coupling constants, can we retain high fidelity in the state transfer process? For example, suppose we try to hop a single qubit one lattice site at a time down a one dimensional chain, ideally performing \begin{equation} 
  \cdots |0\rangle  |\phi\rangle |0\rangle |0\rangle\cdots   \rightarrow \cdots |0\rangle  |0\rangle |\phi\rangle |0\rangle \cdots, \label{eq:hopprotocol}
\end{equation}
which can easily be achieved using local and experimentally realized gates. In the presence of errors, we might obtain (as an example): \begin{equation} 
  \cdots |0\rangle  |\phi\rangle |0\rangle |0\rangle\cdots   \rightarrow  \cdots \left[ \sqrt{1-\epsilon^2} |0\rangle  |0\rangle |\phi\rangle |0\rangle + \epsilon|0\rangle  |\phi\rangle |0\rangle |0\rangle \right] \cdots. \label{eq:naive}
\end{equation}
One might then expect the fidelity $F$ of the transfer process on a chain of length $L$ to scale as \begin{equation}
    F \sim (1-\epsilon^2)^{L-1}, \label{eq:introlowF}
\end{equation}
since some fraction of the wave function is ``lost" at every stage.  In the literature, there have been multiple methods described to overcome this challenge.  For example, one can consider Hamiltonians in one-dimensional chains where qubit transfer between the two edge qubits is protected against noise in couplings \cite{bruderer}.  However, these  approaches are not without their downsides: a significant challenge facing many of these approaches is an asymptotically longer runtime than the naive protocol sketched in (\ref{eq:naive}), when using experimentally realistic couplings or gates.

In this paper, we will show that physical systems with power law interactions provide a natural route to perfoming high fidelity state transfer quickly.  Power law interactions, whereby two qubits $i$ and $j$, located at spatial positions $\mathbf{x}_i$ and $\mathbf{x}_j$, interact via \begin{equation}
    \lVert H_{ij} \rVert \le  h_0 \left(\frac{|\mathbf{x}_i-\mathbf{x}_j|}{d}\right)^{-\alpha}. \label{eq:lrdef}
\end{equation}
Here $h_0$ is a finite constant, $H_{ij}$ represents the subset of terms in the Hamiltonian $H$ which act non-trivially on both qubit $i$ and $j$, and $|\mathbf{x}|$ is a suitable measure of distance in the system. Throughout the text we will use the `Manhattan distance', or shortest distance measured along the lattice axes. The factor of $d^\alpha$ is present to simplify subsequent calculations. Typically, $\mathbf{x}_{i,j}$ would represent the physical locations of the qubits in hardware.  Such interactions are ubiquitous in nature. For example, the Coulomb potential obeys $\alpha=1$, and the interaction potential between two electric or magnetic dipoles obeys $\alpha=3$; the latter interaction is common in many cold atom platforms.  Such platforms were first proposed to speed up state transfer in a quantum system in $d$ spatial dimensions when $\alpha<d+1$ \cite{zack};  more recently, it has been understood how to speed up state transfer when $\alpha<2d+1$ \cite{lr_hierarchy,saito}.  However, the fastest protocols presented in the references above rely on intermediate GHZ states, and so are likely very fragile to error.

The purpose of this paper is to detail and expand upon a different state transfer proposal put forth in \cite{lr_hierarchy}.  Employing time-dependent and tunable long-range interactions obeying (\ref{eq:lrdef}), this ideal (noise free) protocol will achieve perfect state transfer faster than the local hopping protocol (\ref{eq:hopprotocol}) once $\alpha<d+1$.  We review the result of \cite{lr_hierarchy} in Section \ref{sec:review}.  However, because this protocol is based on the dynamics of non-interacting quantum particles hopping on a lattice, we will see that it has two valuable properties.  Firstly, in Section \ref{sec:mp protocol}, we will describe how to transfer $m$ qubits a distance $R$ in a runtime $t_m$ which scales as \begin{equation}\label{eq:t_mp}
    t_m \lesssim t_1 \times \sqrt{2^d m} .
\end{equation}   
This asymptotic scaling is reminiscent of the quadratic speedup of, for example, the quantum walk over the classical walk \cite{ambianis}.  Developing these efficient multi-qubit state transfer protocols will aid in the preparation of complex and highly entangled states of metrological value \cite{kitagawa,leroux}.  Secondly, in Section \ref{sec:random error}, we prove that this protocol is remarkably robust to random uncorrelated noise:  the fidelity $F$ of the protocol is finite in the thermodynamic limit: \begin{equation}
    \lim_{R\rightarrow \infty} \mathbb{E}[F] > 0, \label{eq:introhighF}
\end{equation}
where $\mathbb{E}[\cdots]$ denotes the noise average. The origin of this high fidelity is the large amount of quantum constructive interference among the evolving W states, which speeds up the protocol sufficiently quickly that the most dangerous errors arise at initial (and final) stages of the transfer process.  Similar protocols utilizing W states have been developed for rapid quantum state transfer \cite{1906.02662} with highly non-local interactions. In Section \ref{sec:lr error}, we then describe the performance of a similar protocol that uses non-tunable long-range interactions, which we interpret as a modification of the ideal protocol with correlated errors.  In this case,  we find that \begin{equation}
    F \sim R^{-\gamma}, 
\end{equation}
an intermediate result between (\ref{eq:introlowF}) and (\ref{eq:introhighF}), along with an eventual tradeoff between high fidelity and runtime. 

\section{Review of single-qubit state transfer}\label{sec:review}

\begin{figure}[t]
\centering
\includegraphics[width=.95\textwidth]{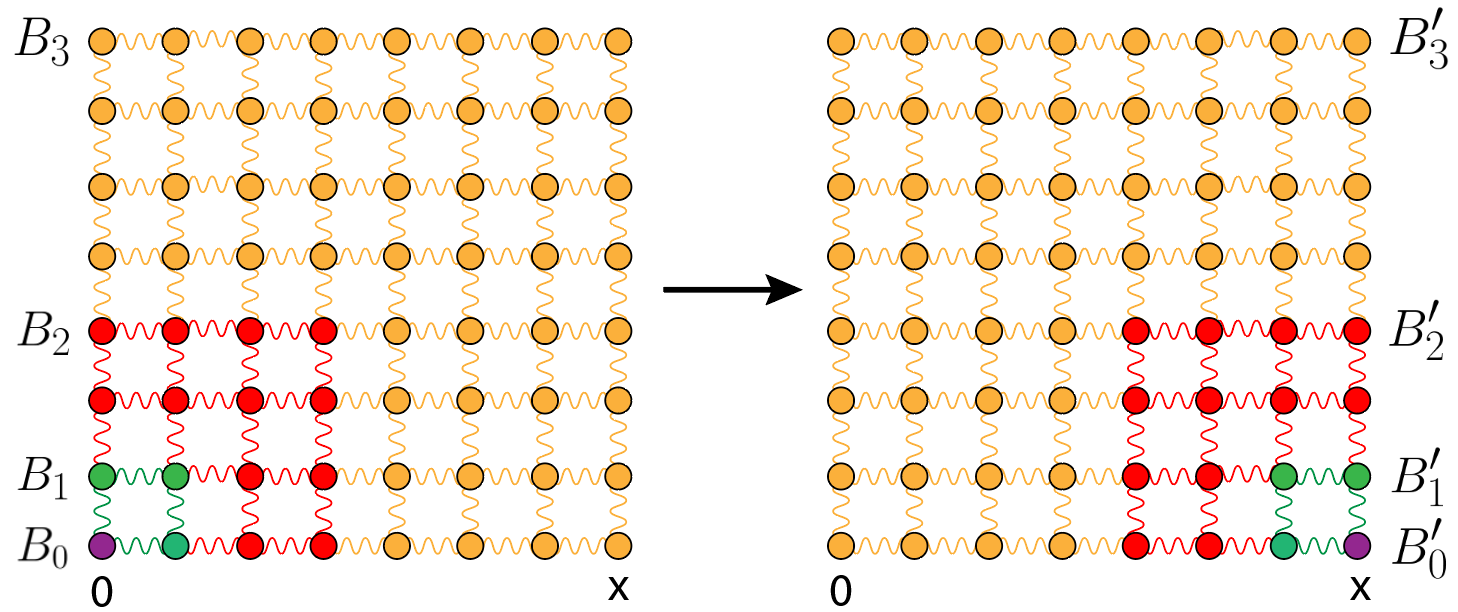}
\caption{A sketch of the ideal single-qubit state transfer protocol in 2D.}
\label{fig:ideal protocol}
\end{figure}

In this section, we review the single-particle state transfer protocol presented in \cite{lr_hierarchy}.  Consider two sites in a $d$-dimensional cubic lattice with lattice spacing unity. For simplicity, we place our two sites along one axis, as shown in Figure \ref{fig:ideal protocol}. We label the original site $\mathbf{0} = (0,0,\ldots,0)$ and the final site $\mathbf{x} = (x,0,\ldots,0)$, and assume the Euclidean distance between the sites is then given by $R=|x|$. The protocol contains two phases: the first phase expands the support from the first site to a uniform superposition of sites containing both the initial and final sites. The second phase collapses this uniform superposition onto the final site. Each phase contains $n\in\mathbb{Z^+}$ steps where
\begin{align}
    n=\left\{\begin{array}{ll}
    \log_2 R & \text { if } R=2^n \\
    \lfloor \log_2R\rfloor+1 & \text { otherwise. }
    \end{array}\right.
\end{align}
For simplicity, let $R$ be a perfect power of $2$ so that we can consider the first case. The second case essentially adds an extra step to both phases of the protocol to correct for the offset, but this can be achieved straightforwardly, as described in \cite{lr_hierarchy}.

\subsection{Expanding phase}
Define a set of cubes $\{B_q\} \in \mathbb{R}^d$, each of (side) length $2^q$ where $q=0,1,\dots ,n$, such that
\begin{align}\label{eq:cubes1}
    \{0\}=B_0 \subset B_1 \subset B_2 \subset \cdots \subset B_n,
\end{align}
and the last cube $B_n$ contains sites $\mathbf{0}$ and $\mathbf{x}$.  The cube $B_n$ is defined by \begin{equation}
    B_n := \lbrace (x_1,\ldots,x_d) : 0\le x_1,\ldots, x_d \le 2^n \rbrace .
\end{equation} Since the length of the final cube is $2^n=R$, the initial and final sites are located on the corners along an edge. To expand the uniform superposition from one ball to the next, we employ the following Hamiltonian:
\begin{align}\label{eq:H_q}
    H_q = \sum_{j \in B_{q-1}} \sum_{k \in \tilde{B}_q} \mathrm{i} h_{jk}(|k\rangle\langle j|-| j\rangle\langle k|),
\end{align}
where $\tilde{B}_q = B_q\setminus B_{q-1}$, and $h_{jk}\in\mathbb{R}$ is a coupling constant for sites between the two regions. In a physical system with long-range couplings, we would have $h_{jk}\sim|j-k|^{-\alpha}$, with $|j-k|$ the distance between the sites $j$ and $k$.  For the ideal protocol, we have a constant coupling factor between the cubes: \begin{equation}
    h_{jk}=C_q=2^{-q\alpha} h_0, \label{eq:idealhjk}
\end{equation}  corresponding to the longest-range (weakest) coupling along the axis containing sites $\mathbf{0}$ and $\mathbf{x}$. For simplicity, we will set $h_0=1$. By permutation symmetry, the wave function takes the form
\begin{align}
    \ket{\psi(t)} = \frac{\cos\theta}{\sqrt{|B_{q-1}|}}\sum_{i\in B_{q-1}}\ket{i} + \frac{\sin\theta}{\sqrt{|\tilde{B}_q|}}\sum_{i\in \tilde{B}_q}\ket{i},
\end{align}
where $|B_q|$ denotes the number of sites in $B_q$, and $\theta = \theta(t)$ is given by
\begin{align}
    \frac{\mathrm{d} \theta}{\mathrm{d} t}&=\frac{\sqrt{|\tilde{B}_q|}}{\cos \theta} \frac{\mathrm{d}\braket{j}{\psi(t)}}{\mathrm{d} t}=-\mathrm{i} \frac{\sqrt{|\tilde{B}_q|}}{\cos \theta}\langle j|H_q| \psi(t)\rangle=C_q \sqrt{|\tilde{B}_q||B_{q-1}|} \\
     \theta(t) &= C_qt\sqrt{|B_{q-1}||\tilde{B}_q|} + \theta_0.
\end{align}
For the initial condition we have $\theta_0=0$, corresponding to a uniform superposition in $B_{q-1}$ and no support inside $\tilde{B}_q$. For a uniform superposition inside $B_q$, we require
\begin{align}\label{eq:t_q}
    \frac{\cos\theta}{\sqrt{|B_{q-1}|}} &= \frac{\sin\theta}{\sqrt{|\tilde{B}_q|}} \quad\implies\quad \theta(t_q) = \tan^{-1}\left(\frac{|\tilde{B}_q|}{|B_{q-1}|}\right) \\
     t_q &= \frac{1}{C_q\sqrt{|B_{q-1}||\tilde{B}_q|}}\tan^{-1}\left(\frac{|\tilde{B}_q|}{|B_{q-1}|}\right).
\end{align}
Thus, if we evolve our wave function with times $\{t_1,t_2,\dots,t_n\}$ corresponding to Hamiltonians $\{H_1,H_2,\dots,H_n\}$ given in Eq.(\ref{eq:H_q}), our state will expand from the initial site $\textbf{0}$ to a large uniform superposition containing both sites $\textbf{0}$ and $\textbf{x}$.

\subsection{Collapsing phase}
For this second phase, we define a similar set of cubes as Eq.(\ref{eq:cubes1}) but around the final site $x$:
\begin{align}
    \{x\}=B'_0 \subset B'_1 \subset B'_2 \subset \cdots \subset B'_n=B_n.
\end{align}
This time however, we want to shrink the uniform superposition to smaller and smaller cubes. This procedure can be accomplished via Hamiltonians analogous to Eq.(\ref{eq:H_q}) but with opposite-sign coupling to mimic inverse-time evolution:
\begin{align}
    H'_q = \sum_{j \in B'_{q-1}} \sum_{k \in \tilde{B}'_q} \mathrm{i} (-h_{jk})(|k\rangle\langle j|-| j\rangle\langle k|).
\end{align}
The time step per Hamiltonian $t'_q$ is analogous to Eq.(\ref{eq:t_q}) but with primes on the $B$'s.

\subsection{Bounding the total runtime}
The total runtime of the protocol $\tau_\mathrm{SP}$ is given by the sum of timesteps in the expanding and collapsing phases:
\begin{align}\label{eq:tau_series}
    \tau_\mathrm{SP} = \sum_{q=1}^n (t_q + t'_q) = 2\sum_{q=1}^n t_q.
\end{align}
To bound the total runtime, we can first begin by bounding each individual timestep $t_q$ in Eq.(\ref{eq:t_q}):
\begin{align}
    t_q = \frac{\phi}{C_q\sqrt{|B_{q-1}||\tilde{B}_q|}} = \frac{2^{q \alpha}\phi}{\sqrt{2^{(q-1) d}\left(2^{q d}-2^{(q-1) d}\right)}}=\frac{2^{d} \phi}{\sqrt{2^{d}-1}} 2^{q(\alpha-d)},
\end{align}
where, for brevity, we define $\phi$ to be \begin{equation}
    \phi := \tan^{-1}\left(|\tilde{B}_q|/|B_{q-1}|\right) = \tan^{-1}(2^d-1).
\end{equation}
For $\alpha \neq d$, the sum in Eq.(\ref{eq:tau_series}) becomes a geometric series which can be readily computed to give the total runtime:
\begin{align}\label{eq:tau_sp}
    \tau_\mathrm{SP} = \frac{2^{d+1}\phi}{\sqrt{2^d-1}} 2^{\alpha-d} \frac{2^{n(\alpha-d)}-1}{2^{\alpha-d}-1} \sim
    \left\{\begin{array}{ll}
    \mathcal{O}(R^{\alpha-d}) & \text { if } \alpha > d \\
    \mathcal{O}(1) & \text { if } \alpha < d.
    \end{array}\right.
\end{align}
For $\alpha=d$ the sum evaluates to
\begin{align}
    \tau_\mathrm{SP} = \frac{2^{d+1}\phi n}{\sqrt{2^d-1}} \sim \mathcal{O}(\log_2 R).
\end{align}

\section{Multi-qubit state transfer}\label{sec:mp protocol}
In the single-particle protocol, we used permutation symmetry to reduce the Hamiltonian
\begin{align}
    H_\mathrm{SP} = \mathrm{i}\sum_{i\in A}\sum_{j\in B} h\left(c_i^\dagger c_j - c_j^\dagger c_i\right)
\end{align}
to the following effective two-level Hamiltonian acting on the distinct regions $A$ and $B$:
\begin{align}
    H_\mathrm{SP}^\mathrm{Eff} = \mathrm{i}C\left(c_A^\dagger c_B - c_B^\dagger c_A\right) = 
    \begin{pmatrix}
    0 & \mathrm{i}C \\
    -\mathrm{i}C & 0
    \end{pmatrix}
\end{align}
where \begin{equation}
    C = h\sqrt{|A||B|}, \label{eq:Cdef}
\end{equation} and 
\begin{subequations}\begin{align}
c_A &= \frac{1}{\sqrt{|A|}} \sum_{i\in A}c_i, \\
c_B &= \frac{1}{\sqrt{|B|}} \sum_{i\in B} c_i 
\end{align}
\end{subequations}
are annihilation operators acting on hybrid single-particle wave functions which are uniform superpositions on all sites.
For transferring multiple particles, we wish to rewrite the effective Hamiltonian into the following form:
\begin{align}\label{eq:H_MP}
    H_\mathrm{MP}^\mathrm{Eff} = \mathrm{i}K\sum_{a=1}^m \left(c_{a,A}^\dagger c_{a,B} - c_{a,B}^\dagger c_{a,A} \right) = \mathrm{i}K
    c^\dagger \begin{pmatrix}
    0 & P \\
    -P' & 0
    \end{pmatrix} c
\end{align}
where $m>1$ is the number of particles we wish to transfer, $K<C$ a constant, and $P, P'$ are $|A|\times |B|$ and $|B|\times |A|$ matrices with $q\leq\min(|A|,|B|)$ ones on the main diagonal, respectively. Our goal is now to construct these $c_{a,B}$ and $c_{a,A}$, making the prefactor $K$ nearly as large as possible, while remaining compatible with (\ref{eq:lrdef}).

\subsection{Orthogonal transformation}
Consider looking for the orthogonal transformations that convert $c_{a,B}$ into $c_j$ (for $j\in B$) and $c_{a,A}$ into $c_i$ (for $i\in A$). We begin our multi-qubit transfer protocol by having all $m$ qubits fully occupy distinct sites (identity basis). The initial states of the qubits are mutually orthogonal, and we wish to transfer them to a set of mutually orthogonal states on the new sites while keeping the couplings as small as possible. We present a simple recursive algorithm to generate these special (non-normalized) orthogonal vectors for dimension $2^w, w\in\mathbb{N}$ in the next paragraph.

For the smallest dimension $(w=1)$, we simply choose the vectors $\mathbf{u}_1 = (1,1)$ and $\mathbf{u}_2 = (1,-1)$. Clearly, $\mathbf{u}_1 \cdot \mathbf{u}_2 = 0$ where the center dot represents the usual dot product for vectors in $\mathbb{R}^2$. To generate half of the next set of vectors $(w=2)$, we simply concatenate the previous vectors onto themselves:
\begin{align}
    \mathbf{v}_1 &= \mathbf{u}_1 \oplus \mathbf{u}_1 \\
    \mathbf{v}_2 &= \mathbf{u}_2 \oplus \mathbf{u}_2 \; .
\end{align}
To generate the second half of the set, we flip the sign on the second vector being concatenated:
\begin{align}
    \mathbf{v}_3 &= \mathbf{u}_1 \oplus -\mathbf{u}_1 \\
    \mathbf{v}_4 &= \mathbf{u}_2 \oplus -\mathbf{u}_2 \; .
\end{align}
In general, let $\mathbf{u}_i^w$ represent the $2^w$-dimensional orthogonal vectors. Then the $2^{w+1}$-dimensional vectors $\mathbf{u}_j^{w+1}$ can be constructed as follows:
\begin{align}
    \mathbf{u}_i^{w+1} &= \mathbf{u}_i^w \oplus \mathbf{u}_i^w \\
    \mathbf{u}_{i+2^w}^{w+1} &= \mathbf{u}_i^w \oplus -\mathbf{u}_i^w \quad,\quad i=1,2,...,2^w \; .
\end{align}
Note that these orthogonal vectors are not normalized:
\begin{equation}
    \sqrt{\mathbf{u}_i^{w}\cdot \mathbf{u}_i^{w}} = 2^{w/2}.
\end{equation}

The next step is to transfer the $m$ qubits onto these orthogonal states. The Hamiltonian takes the form
\begin{align}
    H = \frac{\mathrm{i}K}{2^{w/2}}\begin{pmatrix}
    0 & M_1 \\
    -M_1^{\operatorname{T}} & 0
    \end{pmatrix} \; ,
\end{align}
where $M$ is an orthogonal matrix consisting of the recursively-generated orthogonal vectors mentioned in the previous paragraph: \begin{equation}
    M_1 = 2^{-w/2} \left( \begin{array}{cccc} \mathbf{u}_1^w &\  \mathbf{u}_2^w &\ \cdots &\ \mathbf{u}_{2^w}^w \end{array}\right)
\end{equation}
and the parameter $w$ is chosen to be as small as possible: \begin{equation}
    w = 2^{d\lceil \log_{2^d} m \rceil} < 2^dm.
\end{equation}

For the expanding phase of the protocol, we will concatenate the $M$ matrices horizontally and vertically in order to generate the higher-dimensional block matrix
\begin{align}
    M_{q+1} = \underbrace{\left(\begin{array}{cccc} M_q^{\operatorname{T}} &\ M_q^{\operatorname{T}} &\ \cdots &\ M_q^{\operatorname{T}}
    \end{array}\right)}_{\text{$2^d$ times}} \; .
\end{align}
This sequence of $M$ matrices will cause the qubits to switch between the identity basis and the special orthogonal basis within each $m$-sized block. The method for the collapsing phase will work in a similar manner except in reverse.

The maximum absolute element of our Hamiltonian is given by
\begin{align}
    \max_{i\in A, j\in B} |H_{ij}| = \frac{K}{\sqrt{2^w}} \; .
\end{align}
Hence, the largest value we can choose for $K$ while satisfying (\ref{eq:lrdef}) is
\begin{align}\label{eq:MP coupling}
    K = h\sqrt{2^w} = \frac{C}{\sqrt{2^w}}> \frac{C}{\sqrt{2^d m}}.
\end{align}
with $C$ the single-particle protocol coupling defined in (\ref{eq:Cdef}).

The construction above transfers $m$ qubits in a total time that scales with $\sqrt{m}$. However, for this protocol to work, we need to fully rotate out of every ball $B_q$ at each step. We thus modify the protocol so that all the $B_q$'s are separate from one another, which will only add constant factors to the runtime.

\subsection{Multi-particle runtime}
Let us now carefully evaluate the runtime of this multi-particle transfer protocol. Each timestep runs similarly to that of the single-particle protocol but stretched by a factor of $\sqrt{2^d m}$ and with $\phi=\pi/2$:
\begin{align}
    t_q^\mathrm{MP} < \sqrt{2^dm}\; t_q.
\end{align}
The multi-particle protocol runtime is then bounded by
\begin{align}
    \tau_\mathrm{MP} < 2\sqrt{2^d m}\sum_{q=w}^n t_q = \frac{2^{3d/2}\pi}{\sqrt{2^d-1}} \left(\frac{3}{2}\right)^\alpha \sqrt{m} \sum_{q=w}^n 2^{q(\alpha-d)}
\end{align}
The factor of $\frac{3}{2}$ comes from the fact that the balls $B_q$ are now disjoint.  For $\alpha\neq d$, the sum evaluates to
\begin{align}
    \tau_{\mathrm{MP}}(\alpha\neq d) < \frac{2^{3d/2}\pi}{\sqrt{2^d-1}} \left(\frac{3}{2}\right)^\alpha \sqrt{m} \dfrac{(2R/3+2)^{\alpha-d} - m^{\alpha/d-1}}{2^{\alpha-d}-1} \sim \left\{\begin{array}{ll}
    \mathcal{O} \left( \sqrt{m} \; R^{\alpha-d}\right) & \text { if } \alpha > d \\
    \mathcal{O}(m^{\alpha/d-1/2}) & \text { if } \alpha < d \; .
    \end{array}\right.
\end{align}
For $\alpha=d$, we have
\begin{align}
    \tau_{\mathrm{MP}}(\alpha= d) < \frac{2^{3d/2}\pi}{\sqrt{2^d-1}} \left(\frac{3}{2}\right)^\alpha \sqrt{m} \log_2 \left( \frac{2R/3+2}{m^{1/d}} \right) \sim \mathcal{O}\left( \sqrt{m}\; \log_2 \left(\frac{R}{m^{1/d}}\right) \right)\; .
\end{align}
Since the runtime of middle portion of the protocol scales as $\sqrt{m}$, this portion is faster than simply running the single-particle protocol consecutively for each individual particle (factor of $m$). Thus, we have achieved fast multi-qubit state transfer.

Note that the runtime of this protocol is slower than a simpler nearest neighbor hopping protocol when $\alpha>d+1$ -- however, the high fidelity described in the next section may make this protocol desirable even when $\alpha>d+1$.

\section{Fidelity}
In this section, we describe the fidelity of the ideal protocols described above in the presence of two sources of error.  First, we will describe random noise in the coupling constants, which would arise in a programmable device \cite{Bunyk_2014} with noisy coupling constants.  Second, we will describe the highly correlated errors that can arise when using physical long-range interactions which genuinely depend on the distance between physical qubits in space. For simplicity, we compute the results in the following sections for a single particle $(m=1)$. The many-particle case $(m>1)$ can be easily generalized via (\ref{eq:MP coupling}).

\subsection{Uncorrelated noise in couplings}\label{sec:random error}
Let us begin by treating the case where there is random and uncorrelated error in the value of the coupling constants: namely we have Hamiltonian \begin{equation}
    H = \sum_{j,k} \mathrm{i} h_{jk}(t) \left(c^\dagger_j c_k - c_k^\dagger c_j\right)
\end{equation}  
For a given protocol step (expanding or collapsing), let $Q=2^{qd}$ be the number of sites in $B_q$. The Hamiltonian will contain interaction terms coupling $Q/2^d$ sites in $B_{q-1}$ to the other $Q-Q/2^d$ sites in $\tilde{B}_q=B_q\setminus B_{q-1}$. We introduce an error term so that the overall coupling is
\begin{align}
    h_{jk}(t) \longrightarrow h_{jk}^{(0)}\left( 1 + \epsilon X_{jk} \right) ,
\end{align}
where $h_{jk}^{(0)}$ is the value taken during the ideal protocol, as given in (\ref{eq:idealhjk}), and where $X_{jk}$ are independent and identically distributed random variables chosen from the normal distribution $\mathcal{N}(0,1)$. $\epsilon$ is a tunable parameter characterizing the strength of the disorder. The overall Hamiltonian can thus split into ``ideal" and ``disorder" components:
\begin{align}\label{eq:V_q}
    H_q = H_q^{(0)} + V_q.
\end{align}
We will show that the error \begin{equation}
    \delta_q = \norm{\mathrm{e}^{-\mathrm{i}H_q t_q}-\mathrm{e}^{-\mathrm{i}H_q^{(0)}t_q}}
\end{equation} resulting from this random disorder is strongly bounded; here we use the conventional definition of the operator norm, where $\lVert A\rVert$ represents the maximal singular value of $A$.

In order to bound $\delta_q$, we wish to compare the ``dephasing" rate arising from $V_q$ to the coherent rate of state transfer from $H_q^{(0)}$.  The following result from random matrix theory proves useful (see \cite{mehta} for a review):
\begin{lem}[Bai-Yin's Law \cite{1011.3027}]
    Let $m\in\mathbb{R}$ be a Gaussian random variable with zero mean and variance $\sigma^2$. Let $M$ be an $N_1\times N_2$ matrix whose entries are independent copies of $m$. Without loss of generality, let $N_1\geq N_2$. Then for every $t\geq 0$, there exists a constant $c>0$ such that with probability $1-2\exp (-\frac{1}{2}t^2/\sigma^2)$, we have
    \begin{align}
        \mathbb{E}\norm{M} \leq \sigma ( \sqrt{N_1} + \sqrt{N_2}) + t.
    \end{align}
\end{lem}

Applying Bai-Yin's law to our random disorder Hamiltonian in Eq.(\ref{eq:V_q}), we obtain that for any $\gamma>1$
\begin{align}\label{eq:P_fail}
    \mathbb{P}\left[ \norm{V_q} > \gamma \epsilon C_q \left(  \sqrt{|B_{q-1}|} + \sqrt{|\tilde{B}_q|}\right)\right] < 2\exp[-\frac{1}{2}(\gamma-1)^2 \left(  \sqrt{|B_{q-1}|} + \sqrt{|\tilde{B}_q|}\right)^2].
\end{align}

Now, we seek to bound $\delta_q$. We begin with the Duhamel identity:
    \begin{align}\label{eq:duhamel}
        \mathrm{e}^{(A+B)t} - \mathrm{e}^{At} = \int_0^t \mathrm{d}s \; \mathrm{e}^{(A+B)s}B\mathrm{e}^{A(t-s)}.
    \end{align}
Let $A=-\mathrm{i}H_q^{(0)}$ be the perfect protocol Hamiltonian and $B=-\mathrm{i}V_q$ the random disorder in Eq.(\ref{eq:duhamel}). The error (in 1D) can then be bounded as
\begin{align}\label{eq:applied duhamel}
    \delta_q = \norm{\mathrm{e}^{-\mathrm{i}H_q t_q}-\mathrm{e}^{-\mathrm{i}H_q^{(0)}t_q}} &= \norm{\int_0^{t_q} \mathrm{d}s \; \mathrm{e}^{\mathrm{i}H_q t_q}V_q \mathrm{e}^{-\mathrm{i}H_q^{(0)}(t_q-s)}} \le \norm{V_q}t_q \notag \\
    &\le \gamma \times  \epsilon\phi 2^{d/2}\left[ 1+(2^d-1)^{-1/2}\right] 2^{-qd/2}
\end{align}
We can approximately bound the total error over the entire protocol by summing all the $\delta_q$'s in quadrature:
\begin{align}
    \delta_\mathrm{rand}^2 &\lesssim 2\sum_{q=1}^n \delta_q^2 \leq 2\epsilon^2\phi^2\left[ 1+(2^d-1)^{-1/2} \right]^2 \frac{1-2^{-nd}}{1-2^{-d}} \gamma^2 \notag \\ &\lesssim 2\epsilon^2 \gamma^2 \left[\tan^{-1}(2^d-1)\right]^2 \left[ 1+(2^d-1)^{-1/2} \right]^2 \frac{1-R^{-d}}{1-2^{-d}}  , \label{eq:delta_rand}
\end{align}
If we take the thermodynamic limit $(R\rightarrow\infty)$ we obtain a finite error:
\begin{align}
    \delta_\mathrm{rand}^2(\epsilon,R\rightarrow\infty) \leq \epsilon^2\pi^2, \quad (d=1).
\end{align}
While this argument is not rigorous, a rigorous proof of the protocol's fidelity can be found by simply summing up all the errors linearly:  \begin{equation}
    \delta_\mathrm{rand} \le 2\sum_{q=1}^n \delta_q = 2\epsilon \gamma \tan^{-1}(2^d-1) \left[ 1+(2^d-1)^{-1/2}\right] \frac{1-R^{-d/2}}{1-2^{-d/2}}.
\end{equation}

The probability that this upper-bound is violated can be obtained by summing Eq.(\ref{eq:P_fail}) over all steps of the protocol:
\begin{align}
    P_\mathrm{fail} &< 4\sum_{q=1}^n \exp[-\frac{1}{2}(\gamma-1)^2 2^{qd} \left( 1+2^{1-d}\sqrt{2^d-1} \right)] \\
    &< 4e^{-A} \sum_{q=1}^n 2^{-Aqd} \\
    &< 2^{-A(d+1)+2} \frac{1-R^{-Ad}}{1-2^{-Ad}} \quad,\quad A = \frac{1}{2}(\gamma-1)^2 \left( 1+2^{1-d} \sqrt{2^d-1}\right),
\end{align}
where in the second step we have lower-bounded $2^{qd} < 1+qd\log{2}$ to upper-bound the negative-exponential term, and in the third step we have upper-bounded $e^{-A} \leq 2^{-A}$ since $A\geq 0$. We observe that this probability is finite in the thermodynamic limit $R\rightarrow\infty$.

\begin{figure}[t]
\centering
\includegraphics[width=.45\textwidth]{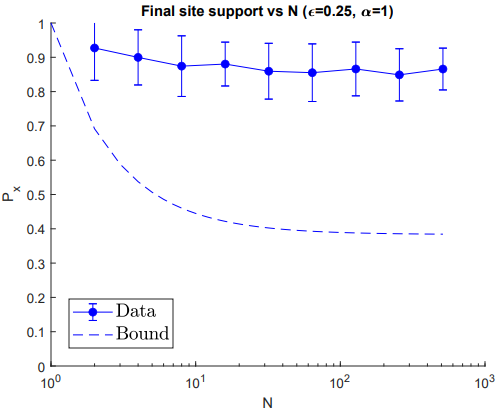}
\includegraphics[width=.45\textwidth]{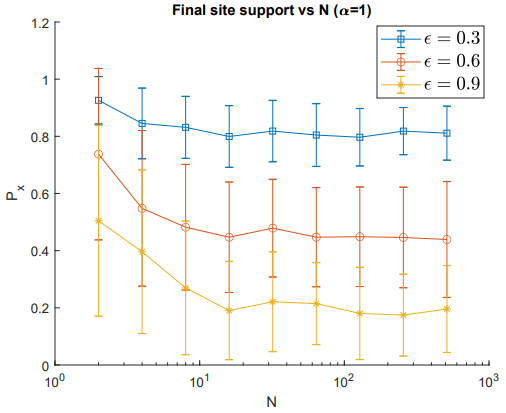}
\caption{The left graph compares the average final site probability for different values of $N$ with the improved $\delta^2$ error bound given by Eq.(\ref{eq:delta_rand}) with $\gamma=1$. The right graph shows the average final site probability after running the full protocol as a function of total number of sites (same as $R$ in 1D) for various values of $\epsilon=0.3,0.6,0.9$ (blue, red, yellow respectively). We average over $100$ iterations of the protocol to obtain the data. Error bars shown are purely statistical.}
\label{fig:random}
\end{figure}

From the numerics shown in Figure \ref{fig:random}, we observe that the final site probability asymptotes to a constant value, as predicted by the argument above. Thus, the state transfer protocol can self-error-correct for random (uncorrelated) errors.  The heuristic bound (\ref{eq:delta_rand}) also reasonably predicts the final fidelity $F=1-\delta^2_{\mathrm{rand}}$. A tighter bound may be obtained by a more careful examination of interference effects in (\ref{eq:applied duhamel}) and in summing the error contributions per protocol step.

We also briefly comment on the possibility of errors arising due to spontaneous emission -- for example, one might consider the $|1\rangle$ qubit to decay by spontaneous emission to the $|0\rangle$ qubit in a cold atomic simulator.  In this case, the fidelity of our algorithm decays exponentially with its runtime: $F\sim \exp[-t]$, since there is always a single $|1\rangle$ qubit somewhere in the W-state (if we are sending this qubit).  However, this is still much better than the fidelity of  GHZ-based protocols due to spontaneous emission, which will scale as $F\sim \exp[-Nt]$ \cite{zack}, where $N$ counts the number of ``active" qubits at any one time.  

\subsection{Physical long-range interactions}\label{sec:lr error}
\begin{figure}[t]
\centering
\includegraphics[width=.95\textwidth]{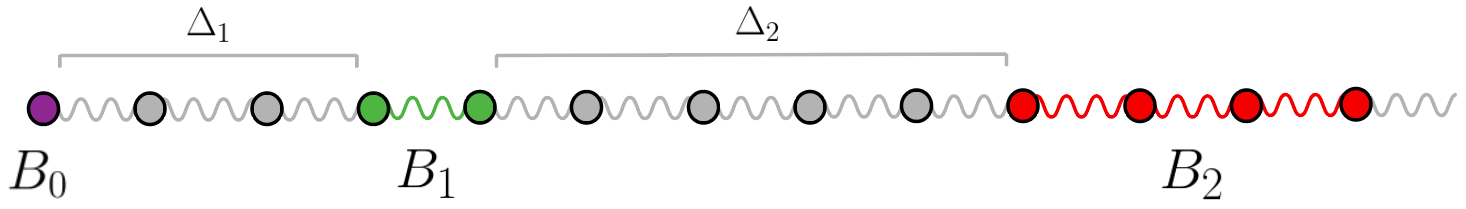}
\caption{The site spacing for $\beta=1$ is shown for the modified long-range protocol.}
\label{fig:gaps}
\end{figure}

In a real system, we may not be able to have constant coupling between sites. Rather, our coupling strength will decay with distance by a power law:
\begin{align}
    h_{jk} = \frac{1}{|j-k|^\alpha},
\end{align}
where $\alpha$ can be a tunable parameter depending on the experimental preparation in trapped ion crystals \cite{exp1}, or is fixed in Rydberg atom arrays \cite{Saffman2010} or dipolar quantum gases \cite{Yan2013}, for example. This kind of correlated error is harmful for our protocol, so to partially mitigate the impact of this error, we modify the protocol by spatially separating the sites as follows: we separate the $B_q$'s such that they no longer overlap and insert a spatial gap of
\begin{align}
    \Delta_q = \beta Q = \beta 2^q
\end{align}
between $B_{q-1}$ and ${B}_q$. Here, $\beta=\beta(\alpha)$ is some prefactor which we can tune based on our final desired probability. Experimentally, this spatial gap can be possibly achieved by ``turning off" certain sites in our lattice. A pictorial representation of the modified protocol is presented in Figure \ref{fig:gaps}.
We also modify the value of $C_q$ to be the middle long-range coupling connecting the centers of $B_{q-1}$ and $B_q$:
\begin{align}
    C_q = \left[ \lceil 2^{q-2}\rceil + \beta 2^q + 2^{q-1}\right]^{-\alpha} = 2^{-q\alpha}\left(\frac{3}{4}+\beta\right)^{-\alpha}.
\end{align}
The runtime during each step of the protocol is then
\begin{align}
    t_q = \frac{\pi}{2C_q\sqrt{|B_{q-1}||B_q}|} = \frac{\pi}{\sqrt{2}}\left(\frac{3}{4}+\beta\right)^\alpha 2^{q(\alpha-1)},
\end{align}
and the total protocol runtime can then be computed via summation. For $\alpha\neq 1$ we have
\begin{align}\label{eq:tau_lr}
    \tau_\mathrm{LR}(\alpha\neq 1) = \pi\frac{2^{\alpha-1/2}}{2^{\alpha-1}-1} \left(\frac{3}{4}+\beta\right)^\alpha \left[\left(\frac{R}{4\beta+3}+1\right)^{\alpha-1} -1\right] \sim
    \left\{\begin{array}{ll}
    \mathcal{O}\left(\left( \frac{3}{4} + \beta\right)R^{\alpha-1}\right) & \text { if } \alpha > 1 \\
    \mathcal{O}\left(\left(\frac{3}{4}+\beta\right)^\alpha\right) & \text { if } \alpha < 1.
    \end{array}\right.
\end{align}
For $\alpha = 1$ we have
\begin{align}
    \tau_\mathrm{LR}(\alpha=1) = \pi\sqrt{2}\left(\frac{3}{4}+\beta\right) \log_2\left(\frac{R}{4\beta+3}+1\right) \sim \mathcal{O}\left(\left(\frac{3}{4} + \beta\right) \log_2 R\right).
\end{align}

In order to bound the long-range error, we split our coupling strength into the ideal and ``error" terms similar to Eq.(\ref{eq:V_q}):
\begin{align}\label{eq:h_LR}
    h_{jk} = \underbrace{\frac{1}{\left((3/4+\beta)2^q\right)^\alpha}}_\text{Ideal} + \underbrace{\frac{1}{|j-k|^\alpha} - \frac{1}{\left((3/4+\beta)2^q\right)^\alpha}}_\text{Error}.
\end{align}
We can upper-bound the positional-dependent first term in the error part by its maximum corresponding to the closest sites between $B_{q-1}$ and ${B}_q$.  For simplicity, we have dropped the $+1$ inside the parenthesis.
\begin{align}
    \frac{1}{|j-k|} \leq \frac{1}{(\beta 2^q)^\alpha}.
\end{align}
The maximum error contribution to $h_{jk}$ per protocol step can then be upper-bounded by
\begin{align}\label{eq:h_q err}
    \max_{j\in B_{q-1},k\in B_q} h_{jk}^\mathrm{err} \leq \frac{1}{2^{q\alpha}}\left(\frac{1}{\beta^\alpha} - \frac{1}{(3/4+\beta)^\alpha}\right) := h_q^\mathrm{max}.
\end{align}

Our error Hamiltonian now has the following block form (per protocol step $q$):
\begin{align}
    H_\mathrm{err} =
    \begin{pmatrix}
        0 & A \\
        A^\dagger & 0
    \end{pmatrix},
\end{align}
\begin{figure}[t]
\centering
\includegraphics[width=.5\textwidth]{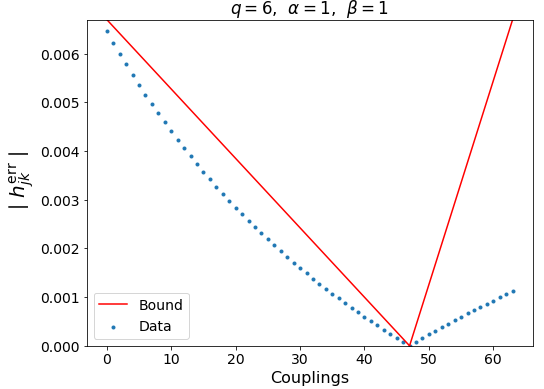}
\caption{A sample plot of the absolute error in (\ref{eq:h_LR}) with the averaged upper-bound for the $q=6$ step of the protocol is shown.}
\label{fig:LR_err}
\end{figure}
where $A$ is a $2^{q-1} \times 2^q$ matrix. The largest singular value of $A$, $\norm{A}_2$, is defined as the largest eigenvalue of $\sqrt{A^\dagger A}$. We will first upper-bound $\norm{A}_1$, or the maximum absolute column sum. Since power-law decay is a convex function, we can upper-bound this sum by replacing each entry of $A$ with the mean value of $h_q^\mathrm{max}/2$. We can do the same with the maximum absolute row sum to upper-bound $\norm{A}_\infty$. A sketch of the bound is presented in Figure \ref{fig:LR_err}. Finally, we invoke the following inequality for matrix norms \cite{golub}:
\begin{align}
    \sigma_\mathrm{max} = \norm{A}_2 \leq \sqrt{\norm{A}_1\norm{A}_\infty}.
\end{align}
Thus, we can upper-bound $\sigma_\mathrm{max}$ of $A$ as follows:
\begin{align}\label{eq:sigma_max}
    \sigma_\mathrm{max} \leq \sqrt{2^{q-1}2^q}\; h_q^\mathrm{max}/2 = 2^{q-2}\sqrt{2}\; h_q^\mathrm{max}.
\end{align}
Let the singular value decomposition of $A=U\Sigma V^\dagger$, where $U,V$ are unitary matrices and $\Sigma$ a diagonal matrix with the (real) singular values. Define the vector $\mathbf{w}$ as
\begin{align}
    \mathbf{w} = 
    \begin{pmatrix}
        \mathbf{u}_j \\
        \mathbf{v}_j
    \end{pmatrix},
\end{align}
where $\mathbf{u}_j,\mathbf{v}_j$ are the respective $j^\mathrm{th}$ columns of $U,V$. After multiplying the left-hand side of $\mathbf{w}$ by $H_\mathrm{err}$ we obtain
\begin{align}
    H_\mathrm{err}\mathbf{w} = 
    \begin{pmatrix}
        0 & U\Sigma V^\dagger \\
        V\Sigma U^\dagger & 0
    \end{pmatrix}\begin{pmatrix}\mathbf{u}_j \\ \mathbf{v}_j \end{pmatrix} = 
    \begin{pmatrix} \sigma_j\mathbf{u}_j \\ \sigma_j\mathbf{v}_j \end{pmatrix} = \sigma_j \mathbf{w}.
\end{align}
Thus, the singular values of $A$ are the eigenvalues of $H_\mathrm{err}$. We can then bound the operator norm of the error Hamiltonian by combining (\ref{eq:h_q err}) and  (\ref{eq:sigma_max}):
\begin{align}
    \norm{H_\mathrm{err}} \leq \sqrt{2} \frac{2^{q-2}}{2^{q\alpha}}\left(\frac{1}{\beta^\alpha} - \frac{1}{(3/4+\beta)^\alpha}\right).
\end{align}
The error at each step of the protocol can then be bounded as
\begin{align}
    \delta_q \leq \norm{H_\mathrm{err}}t_q^\mathrm{LR} = \frac{\pi}{4}\left[\left(1+\frac{3}{4\beta}\right)^\alpha - 1\right],
\end{align}
and the total error can hence be bounded as
\begin{align}
    \delta_\mathrm{LR}^2 &\leq 2\sum_{q=1}^n \delta_q^2 = \frac{\pi^2 n}{8}\left[\left(1+\frac{3}{4\beta}\right)^\alpha - 1\right]^2
\end{align}
which leads to
\begin{align}
    \delta_\mathrm{LR}^2(R,\beta) &\leq \frac{\pi^2}{8}\left[\left(1+\frac{3}{4\beta}\right)^\alpha - 1\right]^2 \log_2 \left( \frac{R}{4\beta+3}+1\right) \quad (d=1).
\end{align}
For $\beta \gg 1$ and $R \gg 1$, we have the approximate result to leading order in $1/\beta$:
\begin{align}
    \delta_\mathrm{LR}^2(R,\beta\gg 1) \lesssim \frac{9\pi^2}{128}\left(\frac{\alpha}{\beta}\right)^2\log_2 \left(\frac{R}{4\beta+3}\right) + \mathcal{O}(\beta^{-4}).
\end{align}
In principle, we can tune $\beta = \beta(\alpha)$ to achieve our desired bound in error.
\begin{figure}[t]
\centering
\includegraphics[width=.45\textwidth]{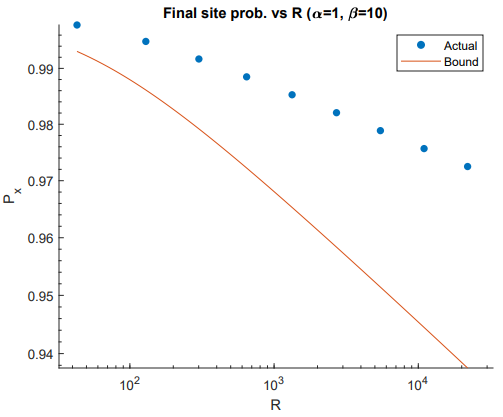}
\includegraphics[width=.45\textwidth]{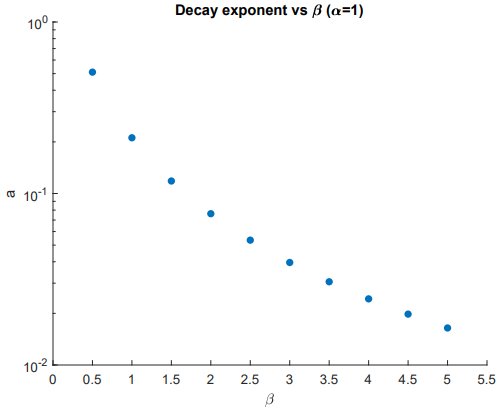}
\caption{The first plot shows the final site probability vs total distance $R$ for the modified protocol with physical long-range interactions. The second plot shows the relationship between the decay exponent $a$ and the gap parameter $\beta$ for $\alpha=1$. The uncertainties from the linear regression are too small to display.}
\label{fig:LR}
\end{figure}
We know that this logarithmic bound cannot be tight since the error cannot exceed 1.   Numerically, we observe a linear correlation between $\log P_x$ and $\log R$ for sufficiently large $n$, implying that the final probability decays as an inverse power-law:
\begin{align}
    \log P_x \sim a\log R + b \implies P_x \sim \mathcal{O}\left( R^{-a}\right),
\end{align}
where $a = a(\alpha,\beta)$ is the decay exponent and can be determined numerically for given parameters $\alpha$ and $\beta$. We run a standard linear regression for several values of $\beta$ and plot the results in Fig. \ref{fig:LR}.

\subsubsection{Fidelity and speed trade-off}\label{sec:fidelity_vs_speed}
Define the fidelity $F$ as our desired probability of measuring the particle at the final site. From our modified long-range protocol, we know that we can increase the final site support at the cost of runtime. So in order to achieve a certain $F$, we can run a fast ($\beta\sim 1$) but inaccurate protocol many times, or we can run a slow ($\beta\gg 1$) but accurate protocol a few times.

Define the effective runtime $\tau_\mathrm{eff}$ as
\begin{align}
    \tau_\mathrm{eff}(F) = \ell \tau_\mathrm{LR}
\end{align}
where $\tau_\mathrm{LR}$ is given in Eq.(\ref{eq:tau_lr}), and
\begin{align}
    \ell = \min_{s}\left[ F > 1-(1-P_x)^s \right]
\end{align}
is the minimum number of times to run the protocol to achieve fidelity $F$. Let $\tau_0 = \tau_\mathrm{LR}(\beta=0)$ denote the runtime of the long-range protocol with no gaps, and let $\tau^* = \tau_\mathrm{eff}/\tau_0$ denote the effective time as a fraction of the gapless runtime.
\begin{figure}[t]
\centering
\includegraphics[width=.45\textwidth]{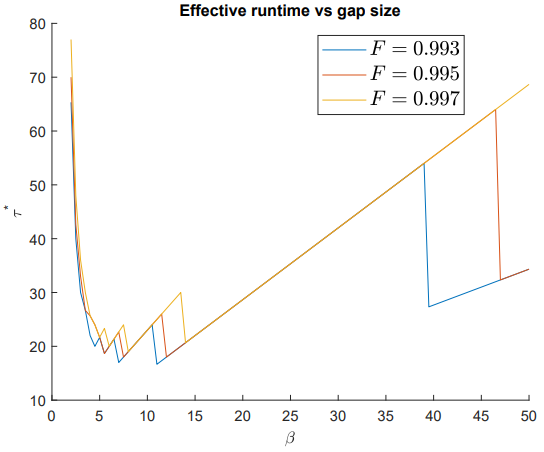}
\includegraphics[width=.45\textwidth]{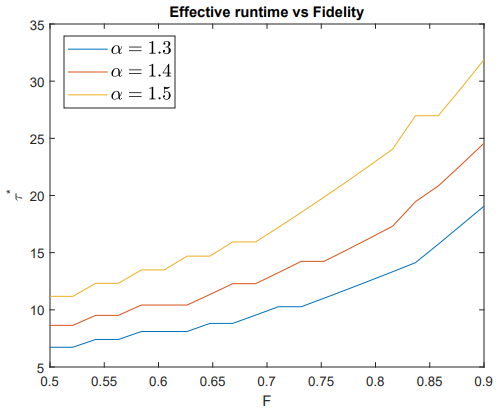}
\caption{The first plot is the effective time fraction $\tau^*$ as a function of the gap size $\beta$ for various values of fidelity $F$ ($\alpha=1$). For a wide range of values of $F$, the minimum value of $\tau^*$ was chosen to produce the second plot.}
\label{fig:tradeoff}
\end{figure}
Figure \ref{fig:tradeoff} shows some ultimate compromise between a large effective fidelity $F$ and the run time of the protocol.   

\section{Conclusion}
We have analyzed in some detail a new protocol for high fidelity state transfer using long-range interactions, and based on intermediate W-states.  Our protocol is effectively immune to uncorrelated errors in programmable couplings, and may be well suited for near-term noisy devices.

One strategy for improving the algorithm's performance with physical long-range interactions could be to ``strobe" the interactions on and off with time, so as to effectively reduce the interaction.  Whether this (or any other method) serves to better mimic the high-fidelity ideal protocol is an interesting problem in quantum engineering of near-term platforms.

A critical property of our protocol is that \emph{all pairs} of qubits are interacting as a consequence of the long-range, power law interactions.   This allows for the extreme quantum coherence that renders finite fidelity in the thermodynamic limit. It is an important open question to understand whether this extremely high fidelity persists on other kinds of quantum hardware, such as combinations of trapped ion crystals interfaced with photons \cite{Monroe_2014,Lekitsch_2017}, which may be more tunable but have more restrictive interaction graphs.

\section*{Acknowledgements}
We thank Andrew Guo and Minh Tran for useful feedback.  AL is supported by a Research Fellowship from the Alfred P. Sloan Foundation.

\bibliography{thebib.bib}
\end{document}